# Field-Free Switching of Perpendicular Magnetic Tunnel Junction via Voltage-Gated Spin Hall Effect for Low-Power Spintronic Memory

Shouzhong Peng, Xiang Li, Wang Kang, He Zhang, Lezhi Wang, Zilu Wang, Zhaohao Wang, Youguang Zhang, Kang L. Wang, *Fellow IEEE*, and Weisheng Zhao, *Senior Member, IEEE*

*Abstract*—Spin Hall effect (SHE) and voltage-controlled magnetic anisotropy (VCMA) are two promising methods for low-power electrical manipulation of magnetization. Recently, magnetic field-free switching of perpendicular magnetization through SHE has been reported with the aid of an exchange bias from an antiferromagnetic IrMn layer. In this letter, we experimentally demonstrate that the IrMn/CoFeB/MgO structure exhibits a VCMA effect of 39 fJ/Vm, which is comparable to that of the Ta/CoFeB/MgO structure. Magnetization dynamics under a combination of the SHE and VCMA are modeled and simulated. It is found that, by applying a voltage of 1.5 V, the critical SHE switching current can be decreased by 10 times owing to the VCMA effect, leading to low-power operations. Furthermore, a high-density spintronic memory structure can be built with multiple magnetic tunnel junctions (MTJs) located on a single IrMn strip. Through hybrid CMOS/MTJ simulations, we demonstrate that fast-speed write operations can be achieved with power consumption of only 8.5 fJ/bit. These findings reveal the possibility to realize high-density and low-power spintronic memory manipulated by voltage-gated SHE.

## I. Introduction

Magnetic random access memory (MRAM) is becoming a promising working memory due to its non-volatility, good endurance, and low power consumption [1], [2]. One of the key challenges for MRAMs is to efficiently manipulate the magnetization of the free layer in the memory cell. Recent studies have demonstrated that data writing can be achieved by spin Hall effect (SHE), in which an in-plane charge current flowing in a nonmagnetic layer generates a vertical spin current, leading to the magnetization reversal of the adjacent magnetic layer [3]–[5]. Voltage-controlled magnetic anisotropy (VCMA) is another promising method for low-power write operations [6]–[12]. It enables us to reduce the interfacial perpendicular magnetic anisotropy (PMA) by applying a voltage across the magnetic tunnel junction (MTJ), thus lowering or even eliminating the energy barrier during magnetization switching.

With a combination of the SHE and VCMA effect, it is feasible to decrease the critical switching current for in-plane MTJs (iMTJs) by using a gate voltage [13]–[15]. For example, Inokuchi *et al.* investigated the switching of iMTJs with SHE currents and control voltages applied to the Ta strips and the MTJ junctions, respectively [14]. By changing the control voltage from +1 V to -1 V, the critical switching current density can be modulated by 3.6 folds.

For memory applications, perpendicular MTJs (pMTJs) are preferable than iMTJs due to their better scalability, faster switching, and lower power consumption [1], [16], [17]. However, an external magnetic field is commonly needed for switching of pMTJs with SHE [18]–[22]. Recently, magnetic field-free switching of perpendicular magnetization was observed in the IrMn/CoFeB/MgO structure, where the antiferromagnetic material IrMn provides an exchange bias and the SHE at the same time [23], [24]. Nevertheless, to our best knowledge, the VCMA effect has never been reported in this structure. Consequently, it is essential and of practical significance to study the VCMA effect and the voltage-gated SHE switching in the IrMn/CoFeB/MgO-based pMTJs.

In this letter, we first deposit IrMn/CoFeB/MgO films and investigate the VCMA effect. A VCMA coefficient of 39 fJ/Vm is obtained in this structure, which is comparable to that of the Ta/CoFeB/MgO structure. Afterwards, magnetic field-free magnetization switching via voltage-gated SHE is presented by solving a modified Landau-Lifshitz-Gilbert (LLG) equation. Finally, a high-density and low-power spintronic memory structure based on this pMTJ is proposed and evaluated by hybrid CMOS/MTJ simulations.

## II. Experimental Investigation of VCMA Effect

The films consisting of IrMn(5nm)/$Co_{40}Fe_{40}B_{20}$(1.05nm)/ MgO(2.5nm)/$Al_2O_3$(5nm) were deposited on thermally oxidized Si substrate using magnetron sputtering, followed by

Manuscript received February 6, 2018. This work was supported by the National Natural Science Foundation of China (Nos. 61571023 and 61627813), the International Collaboration Project B16001, the National Key Technology Program of China (No. 2017ZX01032101), the Academic Excellence Foundation of BUAA for Ph.D. Students, and the National Science Foundation Nanosystems Engineering Research Center for Translational Applications of Nanoscale Multiferroic Systems (TANMS). S. Peng, X. Li, and W. Kang contributed equally to this paper. (*Corresponding author: W. Zhao*)

S. Peng, W. Kang, H. Zhang, L. Wang, Z. Wang, Z. Wang, Y. Zhang, and W. Zhao are with Fert Beijing Institute, BDBC, School of Electronic and Information Engineering, Beihang University, Beijing 100191, China (weisheng.zhao@buaa.edu.cn).
X. Li and K.L. Wang are with Department of Electrical Engineering, University of California, Los Angeles, California 90095, USA.

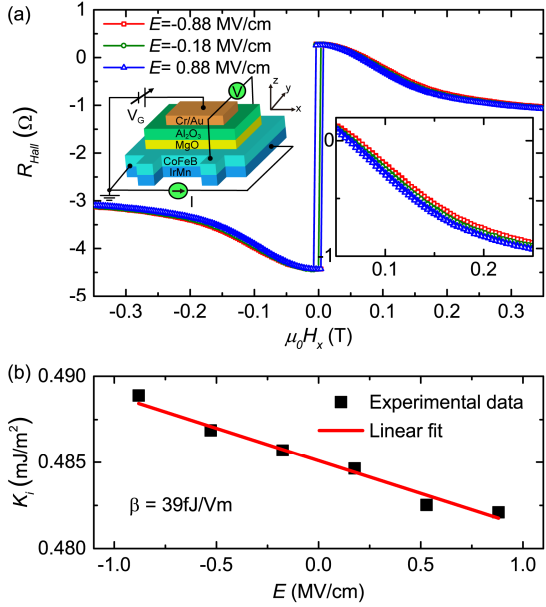

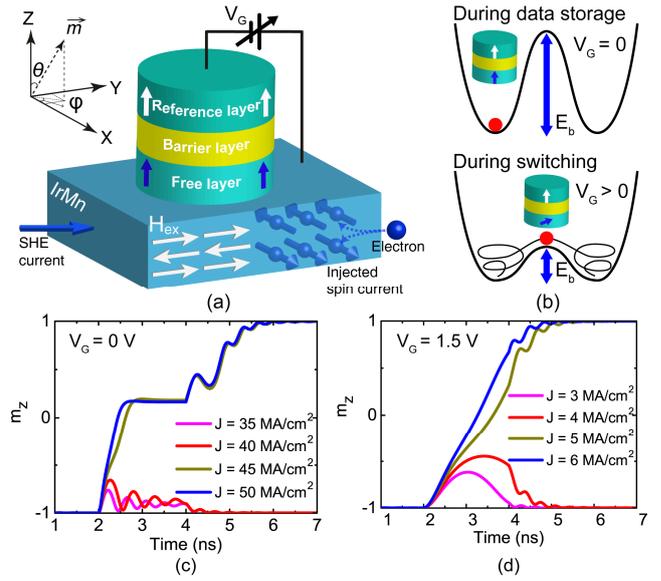

Fig. 1. (a) Anomalous Hall resistance ($R_{Hall}$) of the IrMn/CoFeB/MgO structure under an in-plane magnetic field ($\mu_0 H_x$) and a gate electric field ($E$). The left inset illustrates the Hall bar structure in the measurements. The right inset shows an amplification of the figure with $0.05\text{T} < \mu_0 H_x < 0.25\text{T}$. (b) Interfacial anisotropy constant ($K_i$) as a function of the electric field ($E$) in the MgO layer.

Fig. 2. (a) (b) Schematic of (a) the IrMn/CoFeB/MgO-based pMTJ device and (b) the effect of a gate voltage ($V_G$) on the energy barrier ($E_b$). (c) (d) Perpendicular component of the magnetization ($m_z$) under different SHE currents combined with (c) $V_G = 0$ V and (d) $V_G = 1.5$ V.

annealing at 150℃ for 30 minutes in vacuum (<10$^{-7}$ Torr). Subsequently, the films were patterned into Hall bar devices with dimension of 20 μm × 130 μm. A 33nm-thick Al$_2$O$_3$ layer was then grown on top of the films with atomic layer deposition, covered with a Cr/Au gate electrode. Using a superconducting quantum interference device (SQUID) magnetometer, the saturation magnetization ($M_s$) was measured to be 800 kA/m.

Fig. 1(a) presents the anomalous Hall resistance ($R_{Hall}$) under an in-plane magnetic field. When a gate voltage is applied, the interfacial PMA is changed due to the VCMA effect, leading to variations of the $R_{Hall}$, as clearly shown in the right inset of the Fig. 1(a). Using the method reported in [25], we acquire the electric field dependence of the interfacial anisotropy constant ($K_i$), as shown in Fig. 1(b). The $K_i$ varies linearly with the electric field. From the slope of the linear fit, we obtain a VCMA coefficient ($\beta$) of 39 fJ/Vm for the IrMn/CoFeB/MgO structure, which is comparable to that of the Ta/CoFeB/MgO structure [26], [27].

III. MICROMAGNETIC STUDIES OF VOLTAGE-GATED SHE

Fig. 2(a) illustrates the IrMn/CoFeB/MgO-based pMTJ device, where the lower CoFeB layer is used as the free layer. According to the SHE, an in-plane charge current flowing in the IrMn layer gives rise to a vertical spin current, which exerts spin torque on the free layer. With the aid of an exchange bias ($H_{ex}$) generated by the IrMn layer, no external magnetic field is needed during the SHE-driven switching. In addition, when a gate voltage ($V_G$) is applied across the MTJ, the energy barrier ($E_b$) for magnetization reversal can be reduced due to the VCMA effect, as shown in Fig. 2(b). Consequently, the critical SHE switching current can be modulated by $V_G$.

Magnetization dynamics of the free layer can be described by a modified LLG equation, as

$$\frac{\partial \vec{m}}{\partial t} = -\gamma \mu_0 \vec{m} \times \vec{H}_{eff}(V_G) + \alpha \vec{m} \times \frac{\partial \vec{m}}{\partial t} - \xi \theta_{SH} J_{SHE} \vec{m} \times (\vec{m} \times \vec{\sigma}_{SHE}) \quad (1)$$

where $\vec{m}$ is the magnetization vector of the free layer, $\gamma$ is the gyromagnetic ratio, $\mu_0$ is the vacuum permeability, $\alpha$ is the Gilbert damping constant, $\theta_{SH}$ is the spin Hall angle, $\vec{\sigma}_{SHE}$ is the polarization direction of spin current. $\xi = \gamma \hbar / 2 e t_f M_s$, $\hbar$ is the reduced Planck constant, $e$ is the elementary charge, $t_f$ is the thickness of the free layer. $J_{SHE}$ is the charge current density in the IrMn layer. $\vec{H}_{eff}$ is the effective magnetic field, which includes the exchange bias $\vec{H}_{ex}$, demagnetization field $\vec{H}_{dem}$, and voltage-dependent anisotropy field $\vec{H}_{ani}(V_G)$. It is reported that $H_{ex}$ of 6366 A/m and $\theta_{SH}$ of 0.25 can be obtained simultaneously [28]. The $H_{ani}(V_G)$ is tuned by $V_G$ via the VCMA effect: $H_{ani}(V_G) = 2K_i(V_G)/\mu_0 M_s t_f$, where $K_i(V_G) = K_i(0) - \beta V_G / t_{ox}$ is the interfacial PMA under $V_G$, $t_{ox}$ is the thicknesses of the MgO layer. The $t_{ox}=1.5$ nm, $t_F=1.12$ nm, and MTJ diameter of 80 nm are used in the simulations.

Fig. 2(c) and (d) present the magnetization dynamics when different SHE currents are applied from 2 ns to 4 ns. One can see that, with the aid of the exchange bias from the IrMn layer, magnetic field-free switching can be achieved, in the cost of a large SHE current. Meanwhile, by applying a gate voltage of 1.5 V across the MTJ, the switching current can be dramatically reduced. Fig. 3 shows the critical SHE switching current and the thermal stability factor as a function of the gate voltage. When no gate voltage is applied, the MTJ possesses a thermal stability of 59.8, which is sufficient for reliable data storage during 10 years [1], [29]. In such a case, the critical SHE switching current is 43.2 MA/cm$^2$. When a gate voltage of 1.5 V is applied, the critical switching current drops to 4.5 MA/cm$^2$, which is about 10 times smaller than that under zero gate voltage. These results demonstrate that a gate voltage can be used to effectively modulate the SHE-driven pMTJ switching.

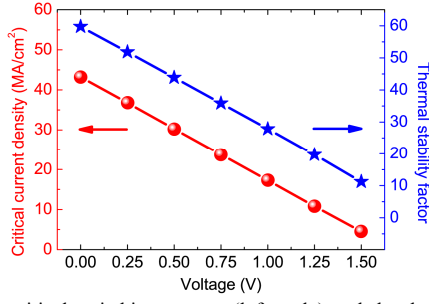

Fig. 3. The critical switching current (left scale) and the thermal stability factor (right scale) as a function of the gate voltage.

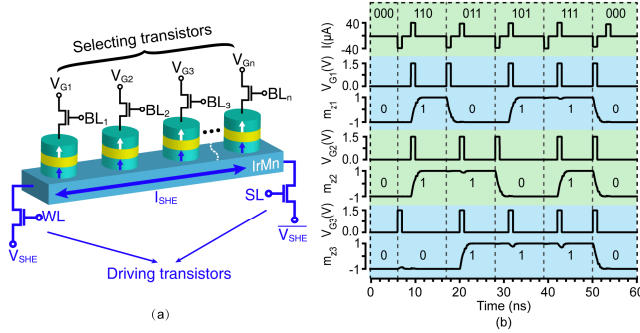

Fig. 4. (a) Schematic of the spintronic memory driven by the voltage-gated SHE. (b) Hybrid CMOS/MTJ simulation waveforms for writing '110', '011', '101', '111' and '000' to three MTJs.

## IV. SPINTRONIC MEMORY DESIGN AND SIMULATION

Using the above-mentioned pMTJ device, we build a spintronic memory with multiple pMTJs located on a single IrMn strip, as shown in Fig. 4(a). During write operations, a bidirectional current is applied in the IrMn strip for the SHE switching, while gate voltages of 1.5 V are employed to the target pMTJs to generate a VCMA effect by activating the selecting transistors. It is worth noticing that all MTJs on a single IrMn strip share two driving transistors. Though each pMTJ requires a selecting transistor, the size of these transistors can be rather small because they do not need to have a large driving capability. As a result, it is promising to achieve higher-density memories than the conventional spin transfer torque (STT) MRAMs and SHE-driven MRAMs [15], [30].

Hybrid CMOS/MTJ simulations are conducted to evaluate the performance of the proposed spintronic memory. An electrical pMTJ model developed with Verilog-A language and a 40 nm CMOS design kit are used in the simulations [31]. Fig. 4(b) presents the simulation results of writing '110', '011', '101', '111' and '000' to three MTJs as an example. The energy dissipation is calculated to be only 8.5 fJ/bit, which is much lower than that of the pure SHE-driven switching. The switching delay is about 2 ns. These results demonstrate that low-power and fast-speed data writing can be achieved in the proposed high-density spintronic memory.

## V. CONCLUSION

In summary, we first experimentally investigated the VCMA effect in the IrMn/CoFeB/MgO structure and obtained a VCMA coefficient of 39 fJ/Vm. Afterwards, micromagnetic simulations were performed to describe the magnetization dynamics of the IrMn/CoFeB/MgO-based pMTJ under the voltage-gated SHE. We found that, by applying a gate voltage of 1.5 V, the critical SHE switching current can be decreased to 4.5 MA/cm$^2$. Finally, a novel high-density spintronic memory structure based on this pMTJ device was proposed and evaluated by hybrid CMOS/MTJ simulations. The results demonstrated that fast-speed write operations can be achieved with power consumption of only 8.5 fJ/bit. Our work will promote the research and development of high-density and low-power spintronic memories.